\documentclass[aps,prl,reprint,superscriptaddress]{revtex4-1}

\usepackage{amssymb,amsmath,mathptmx}
\usepackage{graphicx}
\usepackage{siunitx}
\usepackage{floatrow}
\usepackage{CJK}
\usepackage[breaklinks=true,colorlinks=true,bookmarks=false,urlcolor=blue,
citecolor=blue,linkcolor=blue]{hyperref}
\usepackage{xcolor}

\begin{document}
\title{A Ballistic Graphene Cooper Pair Splitter}

\author{P. Pandey}
\affiliation{Institute of Nanotechnology, Karlsruhe Institute of Technology, D-76021 Karlsruhe, Germany}

\author{R. Danneau}
\email[romain.danneau@kit.edu]{}
\affiliation{Institute for Quantum Materials and Technologies, Karlsruhe Institute of Technology, Karlsruhe D-76021, Germany}


\author{D. Beckmann}
\email[detlef.beckmann@kit.edu]{}
\affiliation{Institute for Quantum Materials and Technologies, Karlsruhe Institute of Technology, Karlsruhe D-76021, Germany}

\begin{abstract}

We report an experimental study of Cooper pair splitting in an encapsulated graphene based multiterminal junction in the ballistic transport regime. Our device consists of two transverse junctions, namely the superconductor/graphene/superconductor and the normal metal/graphene/normal metal junctions. In this case, the electronic transport through one junction can be tuned by an applied bias along the other. We observe clear signatures of Cooper pair splitting in the local as well as nonlocal electronic transport measurements. Our experimental data can be very well described by using a modified Octavio-Tinkham-Blonder-Klapwijk model and a three-terminal beam splitter model. 

\end{abstract}

\maketitle

Superconductors are a source of quantum entangled particles in the form of Cooper pairs \cite{Bardeen1957}. Using a superconductor which is connected to two ferromagnetic or normal metal electrodes, these entangled particles can be separated in space by nonlocal Andreev reflection \cite{Byers1995,Deutscher2000,Falci2001,Deutscher2002,Lesovik2001,Bouchiat2003,Yeyati2007}. In this process, an incoming electron from one electrode is Andreev reflected as a hole in the other electrode while generating a Cooper pair in the superconductor. In the time-reversed process, a Cooper pair from the superconductor is split between the two spatially separated electrodes (Cooper pair splitting, abbreviated CPS). These nonlocal Andreev processes have been experimentally shown in ferromagnet/superconductor/ferromagnet (FSF) \cite{Beckmann2004,Beckmann2007}, and normal metal/superconductor/normal metal (NSN) structures\cite{Russo2005,Zimansky2006,Wei2010} as well as in the semiconductor nanowire based quantum dots \cite{Hofstetter2009,Hofstetter2011,Das2012}, and carbon nanotubes (CNT) and graphene nanoribbons based quantum dots \cite{Herrmann2010,Schindele2012,Schindele2014,Tan2015,Borzenets2016a} coupled to a superconductor. 

Graphene is a system of interest to study nonlocal Andreev processes, firstly, with the theoretical prediction and experimental observation of the specular Andreev reflections in the low energy regime \cite{Beenakker2006,Titov2007,Beenakker2008,Komatsu2012,Efetov2016,Efetov2016a}, and secondly, due to its highly tunable nature \cite{Castro2009,katsnelsonbook}. Single layer graphene is proposed to exhibit nonlocal Andreev processes by taking advantage of its unique electronic band structure and local tuning of the Fermi level \cite{Greenbaum2007,Cayssol2008,Linder2009}. With the improvements in device fabrication techniques \cite{Dean2010,Wang2013}, high quality graphene based superconducting devices have been realized \cite{Calado2015,BenShalom2016,Amet2016,Borzenets2016,Kraft2018,Schmidt2018,Draelos2019}. Nonlocal Andreev processes were experimentally demonstrated in the encapsulated graphene system by exploiting the chiral nature of the edge states in the quantum Hall regime \cite{Clarke2014,Hou2016,Lee2017,Zhang2019}. Recently, CPS has been demonstrated in a vertical double bilayer graphene system \cite{Park2019}. A drawback of CPS via nonlocal Andreev reflection, however, is the exponential suppression of the process as a function of the contact separation on the length scale of the coherence length.

In this letter, we show the experimental observation of CPS in a hexagonal boron nitride (h-BN) encapsulated single layer graphene device which is connected to two superconductors on its parallel edges and to two normal metal electrodes on the other two transverse edges. As a result, two transverse junctions are formed, namely the superconductor/graphene/superconductor (SGS) and the normal metal/graphene/normal metal (NGN) junctions. Due to a difference in the doping density across the two junctions, potential barriers are generated in the graphene channel which act as electronic beam splitters for CPS, similar to the one proposed by Bouchiat \textit{et.\,al.}\,\cite{Bouchiat2003} for the case of single walled carbon nanotube. We employ a modified Octavio-Tinkham-Blonder-Klapwijk (OTBK) model \cite{Blonder1982,Octavio1983,Golubov1995,Perez-Willard2004,Pandey2019} along with a three-terminal beam splitter model to explain our experimental observation. The advantage of this implementation is that the Andreev reflection takes place locally at a single interface, and is not limited by contact spacing.


\begin{figure}
\includegraphics[width=0.9\textwidth]{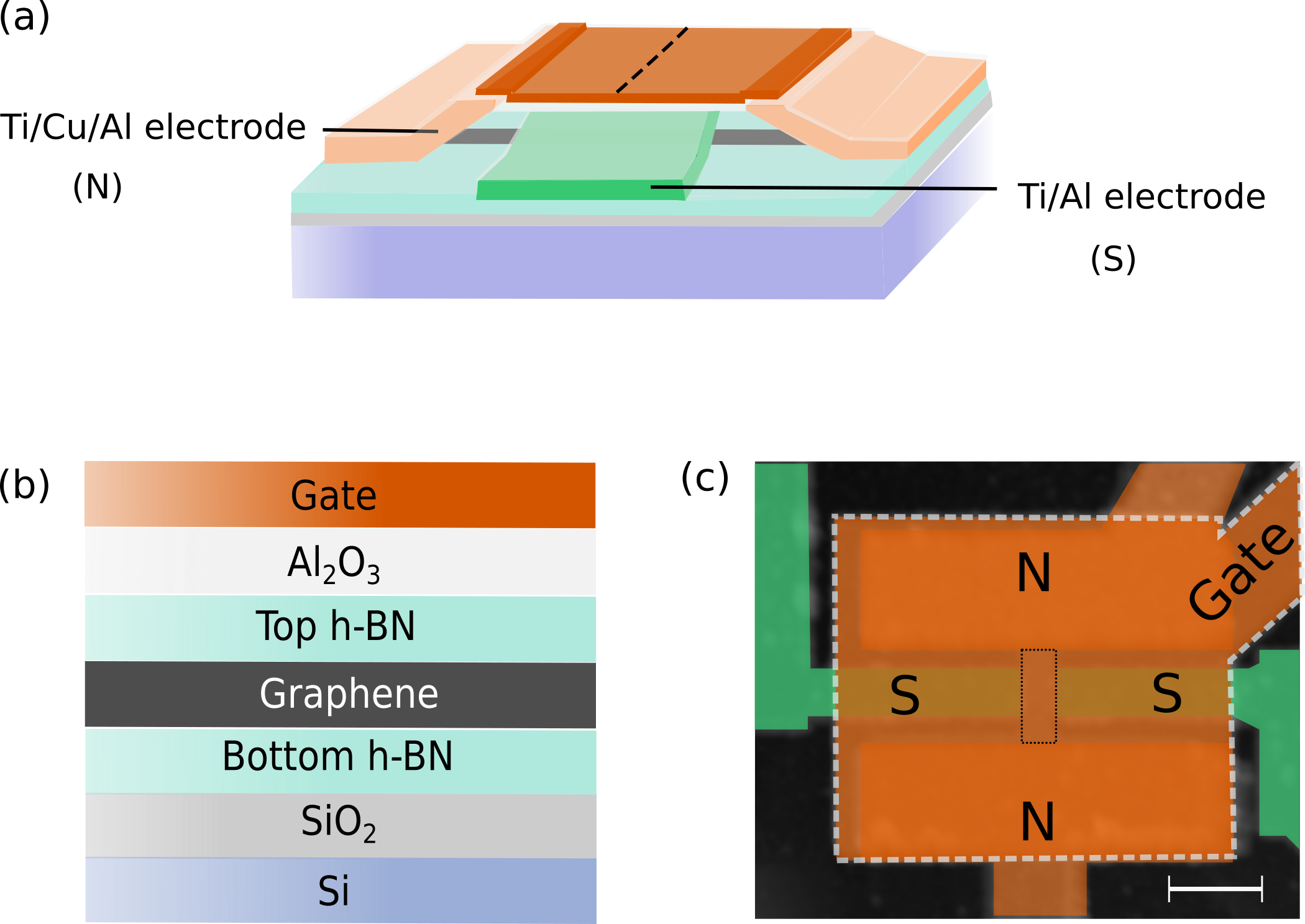}
\caption{(a) Schematic of the device geometry. (b) Cross-section schematic of the device across the dashed line shown in (a). (c) False color atomic force micrograph of the studied device showing the normal (N) and superconducting (S) electrodes. Scale bar is 1\,$\mu$m. The area enclosed by the white dashed lines shows the top gate electrode, and the central area enclosed by the black dotted line shows roughly the h-BN encapsulated graphene.}
\label{fig:device_geo}
\end{figure}

We have used the dry transfer technique for the fabrication of h-BN/graphene/h-BN van der Waals heterostructure which is similar to the one described in Wang \textit{et al.}\,\cite{Wang2013}. Edge contacts to the encapsulated graphene layer were established in a self-aligned manner as described in Kraft \textit{et al.}\,\cite{Kraft2018}, and adapted for two different contact materials in Pandey \textit{et al.}\,\cite{Pandey2019}. Details of
the devices are shown in the Supplemental Material \cite{supmat}. The device schematics are shown in Fig.\,\ref{fig:device_geo}(a) and (b), and the false color atomic force micrograph is shown in Fig.\,\ref{fig:device_geo}(c). The SGS and NGN junctions have the dimensions L=0.36\,$\mu$m and W=0.47\,$\mu$m, and L=1\,$\mu$m and W=0.36\,$\mu$m, respectively, where L is the length of the graphene channel between the electrodes and W is the width of the graphene channel along the electrodes. The sample was mounted in a shielded box attached to the mixing chamber of a dilution refrigerator. Measurement lines were fed through a series of filters to avoid spurious heating by thermal photons or high-frequency interference. Transport measurements were carried out  with the standard low-frequency lock-in detection technique.


\begin{figure}
\includegraphics[width=1\textwidth]{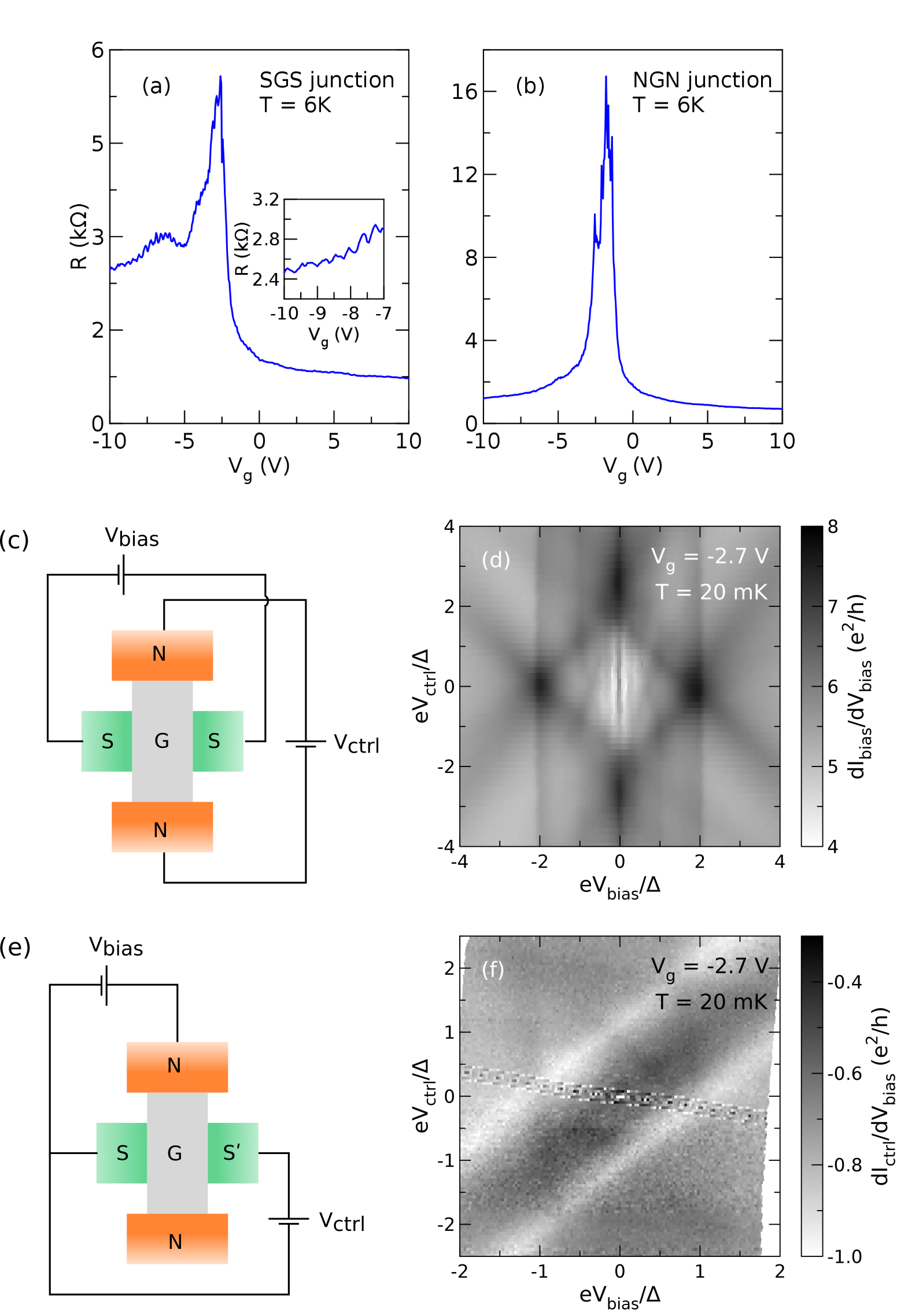}
\caption{Gate dependent resistance of the device across the (a) Superconductor/graphene/superconductor (SGS) junction (inset: Fabry-P\'{e}rot resonances in the p-doped regime), and (b) normal metal/graphene/normal metal (NGN) junction in the normal state. (c) Four-terminal or SGS measurement configuration. (d) Local differential conductance $dI_\mathrm{bias}/dV_\mathrm{bias}$ map across the SGS junction in the superconducting state at $V_\mathrm{g}=-2.7$\,V and $T=20$\,mK under the zero magnetic field in the SGS configuration. (e) Three-terminal or NGS measurement configuration. (f) Nonlocal differential conductance $dI_\mathrm{Ctrl}/dV_\mathrm{bias}$ map across the SGS junction in the superconducting state at $V_\mathrm{g}=-2.7$\,V and $T=20$\,mK without any magnetic field in the NGS measurement configuration.}
\label{fig:exp_data}
\end{figure}

Fig.\,\ref{fig:exp_data}(a) and (b) show the gate dependent resistance of the device across the SGS and NGN junctions, respectively, in the normal state at $T=6$\,K. We observe that the charge neutrality point (CNP) is shifted to the negative gate voltages for both of the junctions which indicates n-type doping of the graphene sheet. The reason for the n-type doping is the charge transfer from the metal contacts which also results in the formation of a potential barrier when the Fermi level of graphene is tuned in the valence band \cite{Giovannetti2008,Khomyakov2009,Khomyakov2010}. Since graphene is doped n-type by the metal contacts, pn-junctions form in the vicinity of the graphene/metal interface when the Fermi level is driven into the hole-doped regime. As a result, an electronic Fabry-P\'{e}rot (FP) cavity is formed which manifests itself in the form of periodic oscillations in the conductance/resistance of graphene, known as the FP resonances, when the charge transport is in the ballistic regime. These resonances can be clearly observed in the p-doped region in the SGS junction as shown in the inset in Fig.\,\ref{fig:exp_data}(a). It clearly indicates that the charge transport in the SGS junction is in the ballistic regime. Additional details on the FP resonances is provided in the Supplemental Material \cite{supmat}. An important point to note is that the two junctions have different doping densities as the CNP for the SGS junction is at the gate voltage $V_\mathrm{g} = -2.6$\,V while the CNP for the NGN junction is at $V_\mathrm{g} = -1.8$\,V. It indicates that the SGS junction is heavily n-doped as compared to the NGN junction. Since the transport across the NGN junction includes the contribution from the SGS junction, it implies that the doping profile along the NGN junction consists of three different regions. In this case, the central graphene region has a different doping density than the two outer regions. It also suggests that apart from the potential barriers at the metal/graphene interface, there are additional potential barriers along the length of the graphene sheet across the NGN junction. It is to be noted that these barriers are expected to be very smooth, in contrast to the gate-defined potential barriers \cite{Young2009,Varlet2014,Du2018}, and therefore, angle dependent transmission due to the Klein tunneling in single layer graphene \cite{Katsnelson2006,Beenakker2008} is expected to modify the transmission in the NGN junction.

We study the Andreev processes in our device in two measurement configurations, namely the four-terminal or SGS and the three-terminal or NGS configurations. Fig.\,\ref{fig:exp_data}(c) shows the SGS measurement configuration where the bias voltage $V_\mathrm{bias}$ is applied across the SGS junction while the control voltage $V_\mathrm{Ctrl}$ is applied across the NGN junction. Fig.\,\ref{fig:exp_data}(d) shows the local differential conductance $dI_\mathrm{bias}/dV_\mathrm{bias}$ map across the SGS junction obtained in the four-terminal configuration at $V_\mathrm{g}=-2.7$\,V (close to the CNP of the SGS junction) and $T=20$\,mK under the zero magnetic field. Note that the $V_\mathrm{bias}$ and $V_\mathrm{Ctrl}$ are normalized with the superconducting gap $\Delta$. The vertical features appearing at $eV_\mathrm{bias}/\Delta = \pm\,1$ and $\pm\,2$ can be assigned to the multiple Andreev reflections in the SGS junction as they appear to be independent of $V_\mathrm{Ctrl}$. However, there are additional conductance features that can be clearly observed in the map. First is the splitted diamond-like pattern that can be observed throughout the entire measurement range. Second are the distinct cross-like features which can be observed in the region $|eV_\mathrm{bias}/\Delta| \leq\,1$ and $|eV_\mathrm{Ctrl}/\Delta| \leq\,2$, and third, a vertical conductance ridge at $|eV_\mathrm{bias}/\Delta| = 0$ when $|eV_\mathrm{Ctrl}/\Delta| \geq\,2$. All of these features are tuned by the $V_\mathrm{bias}$ and $V_\mathrm{Ctrl}$. Given the multiterminal geometry of our device, it suggests that there could be other Andreev processes taking place in the system \cite{Nowak2019}. Next, we employ the three-terminal measurement configuration as shown in Fig.\,\ref{fig:exp_data}(e) where $V_\mathrm{bias}$ is applied across one of the NGS junctions while $V_\mathrm{Ctrl}$ is applied across the SGS junction. The S terminal which is outside the bias circuit and involved only in the control circuit is labeled S$^\prime$ for the sake of clarity. We measure the nonlocal differential conductance $dI_\mathrm{Ctrl}/dV_\mathrm{bias}$ across the SGS junction under the same measurement conditions as for the data shown in Fig.\,\ref{fig:exp_data}(d). The resulting $dI_\mathrm{Ctrl}/dV_\mathrm{bias}$ map is shown in Fig.\,\ref{fig:exp_data}(f). We observe nearly vertical features that could be assigned to the direct transport in the bias circuit and nearly horizontal features that could come from the direct transport in the control circuit. Similar to the data shown in Fig.\,\ref{fig:exp_data}(d), there are other clearly observable conductance features that are influenced by the $V_\mathrm{bias}$ and $V_\mathrm{Ctrl}$, both.


\begin{figure}
\includegraphics[width=1\textwidth]{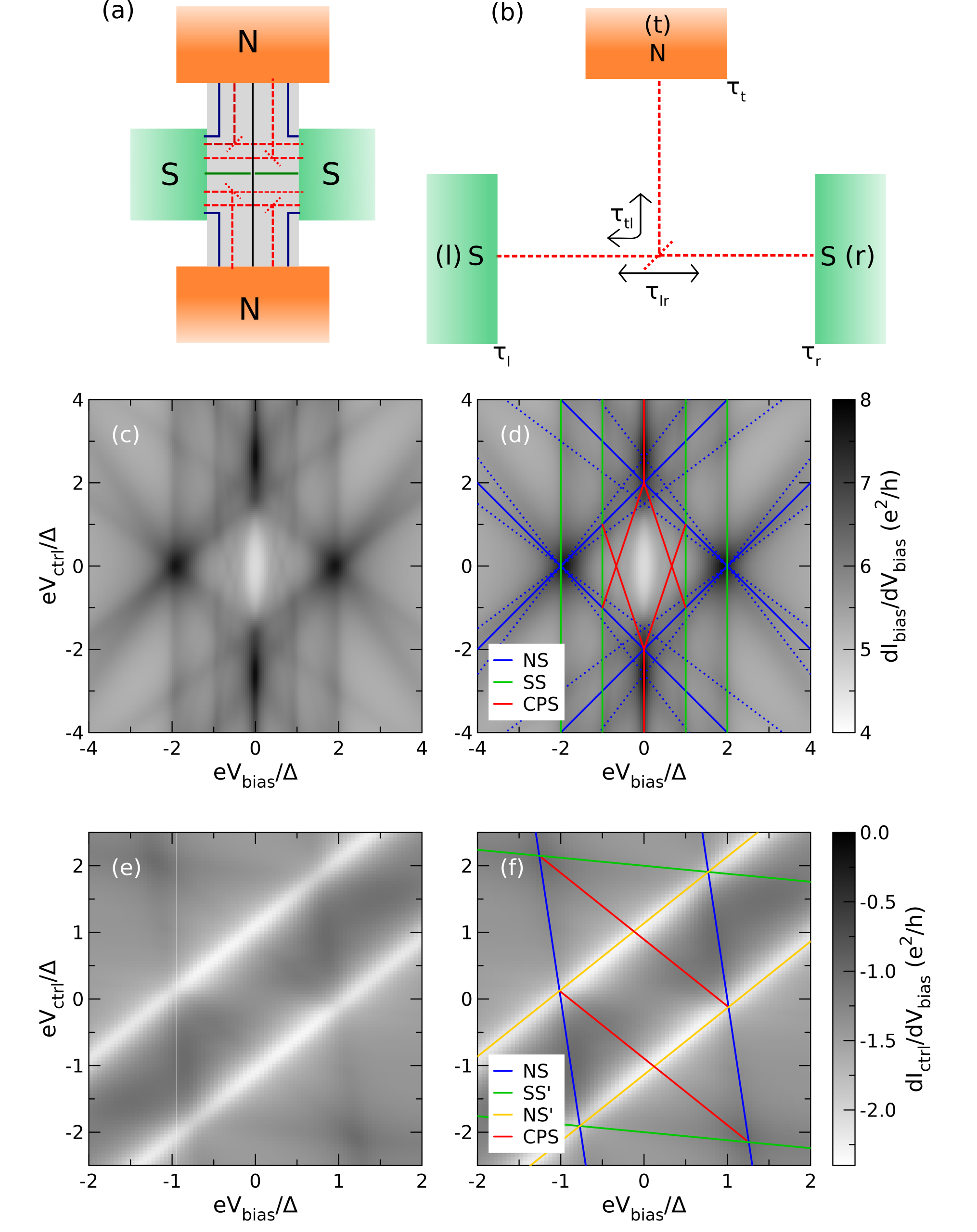}
\caption{(a) Schematics of the model showing local (solid) and nonlocal (dashed) transport channels. (b) Three-terminal beam splitter for the nonlocal transport channels as shown in (a) along with the distribution functions. t, l and r stand for top, left and right, respectively. Differential conductance maps generated with the model: local differential conductance $dI_\mathrm{bias}/dV_\mathrm{bias}$ map in the SGS measurement configuration: (c) without, and (d) with the guidelines, and nonlocal differential conductance $dI_\mathrm{Ctrl}/dV_\mathrm{bias}$ map in the NGS measurement configuration: (e) without, and (f) with the guidelines.}
\label{fig:model_comp}
\end{figure}

In order to interpret the experimental data, we employ a modified Octavio-Tinkham-Blonder-Klapwijk (OTBK) model \cite{Blonder1982,Octavio1983,Golubov1995,Perez-Willard2004,Pandey2019} for the transport between each pair of terminals, as indicated by the solid lines in Fig.~\ref{fig:model_comp}(a). To account for nonlocal transport processes, we include a beam splitter model in addition to the two-terminal contributions. The beam splitters in our device are expected to form due to interfaces between regions of different doping densities where Klein tunneling leads to an angle-dependent partial transmission and reflection \cite{Katsnelson2006,Beenakker2008}. Given the geometry of our device, nonlocal Andreev processes that involve both of the S terminals and one N terminal are most likely to take place. From the pairwise two-terminal resistances measured in the normal state, we know that the device is highly asymmetric with respect to the four corners, especially near the Dirac point. To incorporate nonlocal processes and the asymmetry of the device at the same time, we model four beam splitters which follow the four corners of the device, as indicated by dashed lines in Fig\,\ref{fig:model_comp}(a). As an example, the beam splitter in the top left corner is shown in detail in Fig.\,\ref{fig:model_comp}(b). To follow the corner, an electron from the top N terminal can be transmitted only to the left S terminal, with probability $\tau_\mathrm{tl}$. An electron from the right S terminal can be transmitted with probability $\tau_\mathrm{lr}=1-\tau_\mathrm{tl}$ to the left terminal. For simplicity, we assume that there is no reflection back to the left terminal. It is to be noted that the beam splitters also contribute to the local processes between the NS and SS terminals. Therefore, the model also includes nonlocal MAR processes where an initial nonlocal Andreev process starts a MAR cycle between the two superconductors. The model is described in detail in the Supplemental Material \cite{supmat}.

Fig.\,\ref{fig:model_comp}(c) and (d) show the $dI_\mathrm{bias}/dV_\mathrm{bias}$ maps without and with the guidelines, respectively, as generated with the model for the SGS measurement configuration and corresponding to the experimental data shown in Fig.\,\ref{fig:exp_data}(d). The transmission coefficients at the superconducting terminals were chosen to be 0.72 while they were 0.7 at the normal metal terminals. Comparing Fig.\,\ref{fig:exp_data}(d) with \ref{fig:model_comp}(c), it can be readily seen that the experimental data and the model are in very good agreement. To interpret the observed conductance features, we compare Fig.\,\ref{fig:exp_data}(d) with \ref{fig:model_comp}(d). As expected, the vertical conductance features at $eV_\mathrm{bias}/\Delta = \pm\,1$ and $\pm\,2$ (marked SS in Fig.\,\ref{fig:model_comp}(d)) appear due to the direct transport between the two superconductors. The diamond-like feature (marked NS in Fig.\,\ref{fig:model_comp}(d)) appears because of the transport across the NGS corners of the device. The splitting spread of this feature results from the asymmetry of the corner contacts. Most interesting are the cross-like features which appear in the region $|eV_\mathrm{bias}/\Delta| \leq\,1$ and $|eV_\mathrm{Ctrl}/\Delta| \leq\,2$, and the vertical conductance ridge at $|eV_\mathrm{bias}/\Delta| = 0$ when $|eV_\mathrm{Ctrl}/\Delta| \geq\,2$ (marked CPS in Fig.\,\ref{fig:model_comp}(d)). These features appear as a result of the nonlocal transport involving the N and S terminals. While the cross-like features can be unambiguously assigned to the Cooper pair splitting, the vertical conductance ridge remains slightly ambiguous due to the possibility of a very weak supercurrent flowing through the SGS junction. It is to be noted that the measurements were conducted close to the CNP of the SGS junction where the resistance of the graphene channel is very high, as can be seen in Fig.\,\ref{fig:exp_data}(a). Therefore, the magnitude of the supercurrent is too small to be measured, however, its contribution can not be completely ruled out.

Fig.\,\ref{fig:model_comp}(e) shows the $dI_\mathrm{Ctrl}/dV_\mathrm{bias}$ map for the NGS configuration as generated with the model which corresponds to the experimental data shown in Fig.\,\ref{fig:exp_data}(f), while Fig.\,\ref{fig:model_comp}(f) shows the same map but with the guidelines. Comparing Fig.\,\ref{fig:exp_data}(f) with \ref{fig:model_comp}(e), we can see that there is a qualitative agreement between the experimental data and the model. It is to be noted that the experimental measurement in the NGS configuration was two-probe instead of the pseudo four-probe as in the SGS configuration. As a result, the experimental data includes the series resistances from the filters in the measurement lines. Therefore, the comparison between the experimental data and the model in the NGS configuration is only qualitative. Fig.\,\ref{fig:exp_data}(f) and \ref{fig:model_comp}(f) can be compared for the interpretation of the various conductance features. There are three different sets of features which appear due to the direct transport in the device, namely the nearly vertical ones due to the transport across the bias junction (marked NS in Fig.\,\ref{fig:model_comp}(f)), the nearly horizontal features due to the transport across the control junction (marked SS$^\prime$ in Fig.\,\ref{fig:model_comp}(f)), and the features with negative contribution which appear due to the transport across the NGS$^\prime$ junction  (marked NS$^\prime$ in Fig.\,\ref{fig:model_comp}(f)). In this case too, we clearly observe the conductance features (marked CPS in Fig.\,\ref{fig:model_comp}(f)) which appear only due to the nonlocal Andreev process in the device.


\begin{figure}
\includegraphics[width=1\textwidth]{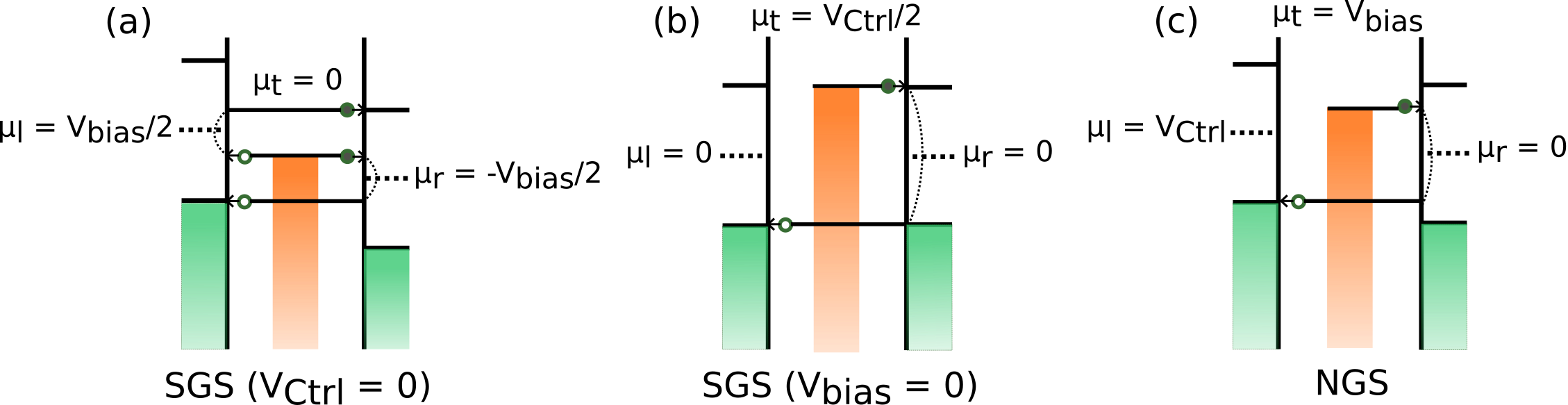}
\caption{Bias conditions for the observation of nonlocal Andreev processes. In the SGS configuration at (a) $V_\mathrm{Ctrl}=0$ and (b) $V_\mathrm{bias}=0$, and (c) in the NGS configuration. $\mu$ represents the chemical potential of the respective terminal. Filled and empty circles denote the electrons and holes, respectively.}
\label{fig:Bias_cond}
\end{figure}

As seen in Fig.\,\ref{fig:exp_data}(d) and (f), the nonlocal Andreev processes appear in a certain $V_\mathrm{bias}$ and $V_\mathrm{Ctrl}$ range. The onset of these processes can be explained by considering the energy thresholds. For the SGS configuration at $V_\mathrm{Ctrl}=0$, the threshold for the onset of nonlocal Andreev process is at $eV_\mathrm{bias}=2\Delta/3$ as shown in Fig\,\ref{fig:Bias_cond}(a). In this case, an electron which incidents from the normal metal is Andreev reflected at the left superconductor while the resulting hole enters the right superconductor. On the other hand, for a hole which incidents from the normal metal 
onto the right superconductor, the Andreev reflected electron enters the left superconductor. As soon as $|V_\mathrm{Ctrl}| > 0$, the threshold for $V_\mathrm{bias}$ changes and this results in the cross-shaped feature observed in Fig.\,\ref{fig:exp_data}(d) in the region $|eV_\mathrm{bias}/\Delta| \leq\,1$ and $|eV_\mathrm{Ctrl}/\Delta| \leq\,2$. For the condition $V_\mathrm{bias}=0$ in the SGS configuration, the nonlocal Andreev reflection is enabled as soon as $eV_\mathrm{Ctrl} \geq 2\Delta$ as illustrated in Fig.\,\ref{fig:Bias_cond}(b). This is seen as the vertical conductance ridge in Fig.\,\ref{fig:exp_data}(d) in the region $|eV_\mathrm{bias}/\Delta| = 0$ when $|eV_\mathrm{Ctrl}/\Delta| \geq\,2$. Due to a peak in the Andreev reflection probability at $\epsilon = \Delta$, a conductance peak is observed when the gap features of the two superconductors are aligned with each other. In the NGS configuration, the nonlocal Andreev process is enabled when $eV_\mathrm{bias} + eV_\mathrm{Ctrl} = \Delta$ as shown in Fig.\,\ref{fig:Bias_cond}(c) and results in the conductance feature observed in Fig.\,\ref{fig:exp_data}(f). 

Our model is agnostic to the actual implementation of the beam splitters in the device. Possible candidates are: 1.~pn junctions between different doping levels across the NGN and SGS junction as evidenced by the different positions of the Dirac points in Figs.~\ref{fig:exp_data}(a) and \ref{fig:exp_data}(b), 2.~the pn-junctions near the superconducting interfaces which give rise to the FP cavity, and 3.~more complicated doping inhomogeneities near the Dirac point as evidenced by the conductance asymmetry in the normal state (see also Supplemental Material). Our conclusions about the bias thresholds for CPS are however independent of this detail.

 
To conclude, we have shown the experimental observation of CPS in a multiterminal graphene device in the ballistic regime. Our device takes advantage of its simple geometry with two different transverse junctions and tunable doping profile across the graphene channel to split the Cooper pair. Furthermore, the signature of CPS is clearly observed in the local as well as nonlocal differential conductance measurements. The experimental observation is very well supported by the modified OTBK model and the three-terminal beam splitter model. Our work shows an experimentally accessible way to achieve and control the spatially separated quantum entangled particles in graphene by using the local tuning of the Fermi level and the angle dependent transmission through the potential barriers.

\acknowledgments{The authors thank R. M\'{e}lin and B. Dou\c cot for fruitful
discussions. This work was partly supported by Helmholtz
society through program STN and the DFG via the projects
DA 1280/3-1. P.P. acknowledges support from Deutscher
Akademischer Austauschdienst (DAAD) scholarship.}

\end{document}


\title{Supplemental Material for \\ ``A Ballistic Graphene Cooper Pair Splitter''}

\author{P. Pandey}
\affiliation{Institute of Nanotechnology, Karlsruhe Institute of Technology, D-76021 Karlsruhe, Germany}

\author{R. Danneau}
\email[{\color{blue}romain.danneau@kit.edu}]{}
\affiliation{Institute for Quantum Materials and Technologies, Karlsruhe Institute of Technology, Karlsruhe D-76021, Germany}


\author{D. Beckmann}
\email[{\color{blue}detlef.beckmann@kit.edu}]{}
\affiliation{Institute for Quantum Materials and Technologies, Karlsruhe Institute of Technology, Karlsruhe D-76021, Germany}

\maketitle

\clearpage

\section{Device fabrication methods}

h-BN crystallites ($\sim$\,28 and $\sim$\,18\,nm thick for the top and bottom h-BN layer, respectively) and single layer graphene were exfoliated from the commercially available h-BN powder (Momentive, grade PT110) and natural graphite (NGS Naturgraphit GmbH), respectively, on Si substrates with 300\,nm thick SiO$_2$ layer. The h-BN/graphene/h-BN heterostructure was fabricated by the dry transfer method similar to the transfer technique shown by Wang \textit{et.\,al.}\,\cite{Wang2013}. The fabricated heterostructure was then transferred on a Si substrate with 1000\,nm thick SiO$_2$ layer. The encapsulated graphene layer was connected to the metal electrodes on its edges in a self-aligned manner as described in Kraft \textit{et al.}\,\cite{Kraft2018}, and adapted for two different contact materials in Pandey \textit{et al.}\,\cite{Pandey2019}. We have used Ti/Al (5 and 60\,nm, respectively) for the superconducting electrodes and Ti/Cu/Al (5, 80 and 5\,nm, respectively) for the normal electrodes. For the fabrication of the top gate electrode, a 25\,nm thick Al$_2$O$_3$ layer was deposited by using atomic layer deposition. It was followed by the metallization for the top gate electrode made of Ti/Cu/Al (5, 100 and 5\,nm, respectively). The Ti layer serves as an adhesive layer for all of the electrodes while the thin Al layer in the normal metal and gate electrodes serves as the capping layer to protect the underlying Cu layer from oxidation.

\section{Ballistic transport}

\begin{figure}
\includegraphics[width=0.8\textwidth]{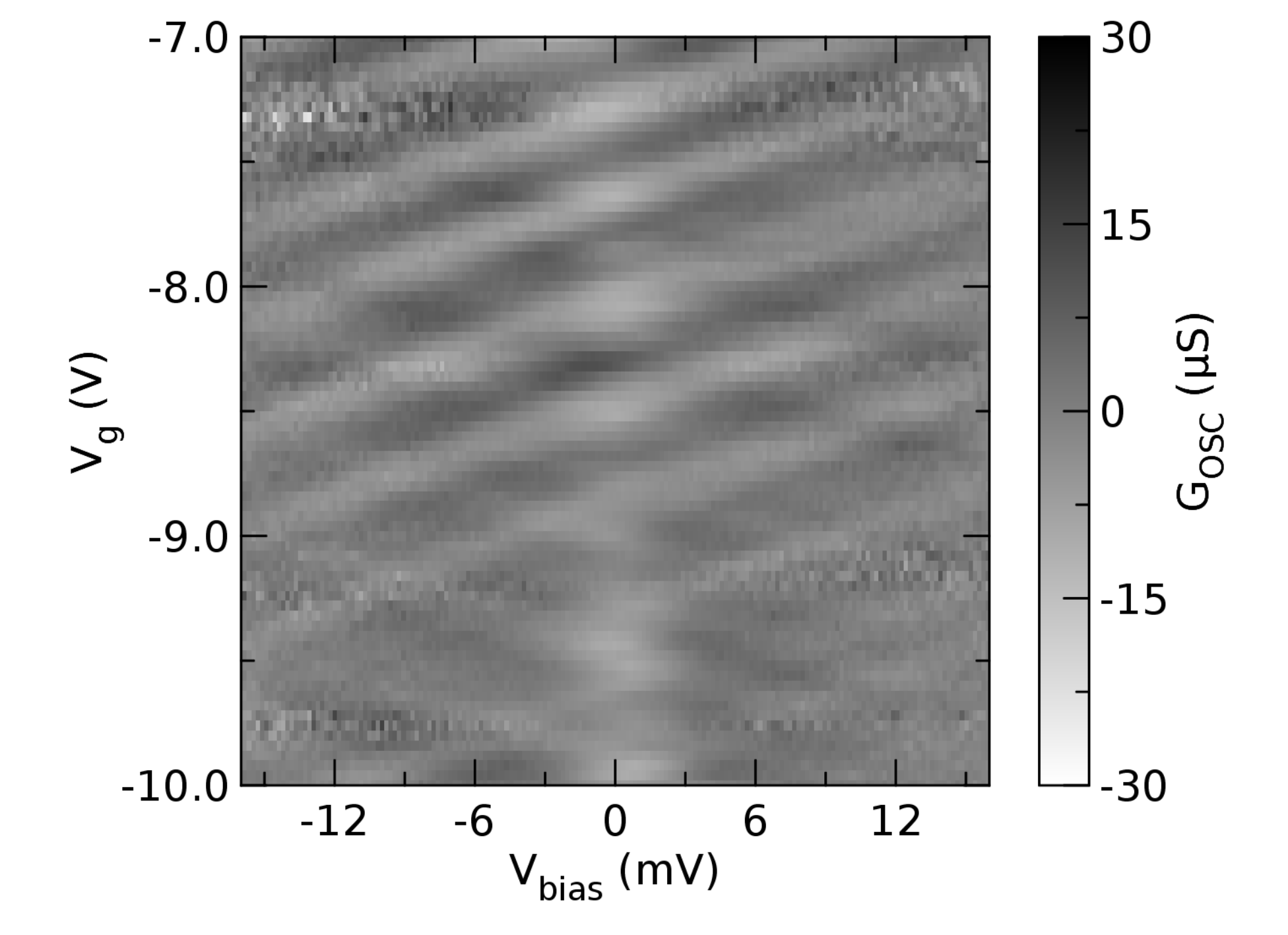}
\caption{Fabry-P\'{e}rot resonances observed as the conductance oscillations across the SGS junction as a function of the gate voltage $V_\mathrm{g}$ and bias voltage $V_\mathrm{bias}$.}
\label{fig:FP_osc}
\end{figure}

Fig.\,\ref{fig:FP_osc} shows the checkerboard conductance pattern indicating the Fabry-P\'{e}rot (FP) resonances across the SGS junction as tuned by the gate voltage $V_\mathrm{g}$ and bias voltage $V_\mathrm{bias}$. Since the conductance of the device varies strongly with $V_\mathrm{g}$ and $V_\mathrm{bias}$, a non-oscillating background conductance was subtracted to enhance the visibility of the conductance oscillations. The length $L_\mathrm{C}$ of the electronic FP cavity can be calculated with the expression $k_\mathrm{f}L_\mathrm{C}= N\pi$ for normal incidence of the charge carriers at the graphene/metal interface, here $k_\mathrm{f}$ is the Fermi wavevector and $N$ is an integer. $k_\mathrm{f}$ can be tuned by $V_\mathrm{g}$ as per the relation $k_\mathrm{f} = \sqrt{n\pi}$, where n is the charge carrier density given as $n= \alpha_\mathrm{g}V_\mathrm{g}$ with $\alpha_\mathrm{g}$ being the gate coupling efficiency. $L_\mathrm{C}$ is found to be 353\,$\pm$\,3\,nm with $\alpha_\mathrm{g} =$\,3.496\,x\,$10^{11}$\,V$^{-1}$cm$^{-2}$ ($\alpha_\mathrm{g}$ is determined with the Shubnikov de Haas oscillations). Since $L_\mathrm{C}$ agrees with the graphene channel length across the SGS junction, it is evident that the observed conductance oscillations are occurring due to the FP interferences where the pn-junctions forming at the graphene/metal interfaces act as the partially transmitting interfaces of the cavity. It serves as a proof that the charge transport across the SGS junction is in the ballistic regime.

\section{Model}

\begin{figure}
\includegraphics[width=1\textwidth]{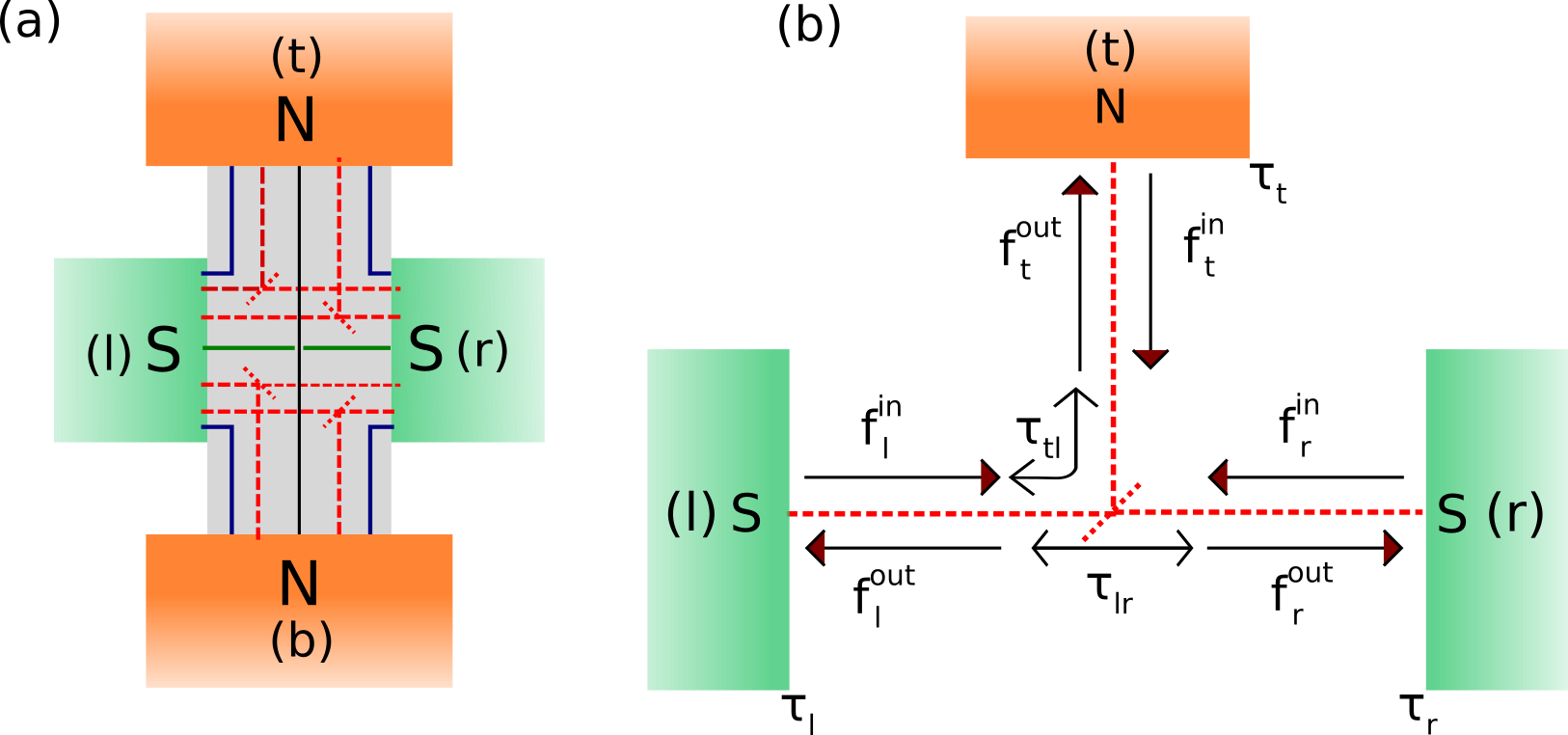}
\caption{(a) Schematics of the model showing the local (solid) and nonlocal (dashed) transport channels.  t, l, r and b stand for top, left, right, and bottom, respectively. (b) Three-terminal beam splitter for the nonlocal transport channels as shown in (a) along with the distribution functions. Note that these are the same figures as Fig.\,3(a) and (b) in the main text.}
\label{fig:CAR_model}
\end{figure}

We model the system with graphene being a multichannel conductor which is connected to four different terminals as shown in Fig.\,\ref{fig:CAR_model}(a), the superconducting (S) terminals on the left and right (labeled $\mathrm{l}$ and $\mathrm{r}$), and the normal metal (N) terminals on the top and bottom (labeled $\mathrm{t}$ and $\mathrm{b}$). These terminals have the transmission probabilities $\tau_\mathrm{l}, \tau_\mathrm{r}, \tau_\mathrm{t}$ and $\tau_\mathrm{b}$, respectively. We account for the local or direct transport processes by considering that the terminals are connected with ideal conductors (shown as solid lines in Fig.\,\ref{fig:CAR_model}(a)), while the nonlocal transport processes are addressed by employing the three-terminal beam splitters (shown with dashed lines in Fig.\,\ref{fig:CAR_model}(a) and (b)). The number of transport channels for each of the conductors are accounted by the parametric conductance $G_\mathrm{\alpha\,\beta}$ where $\alpha$, $\beta$ denote the terminals. It results in six independent conductances that allows us to model the device in a realistic sense by taking into account the asymmetry between the various graphene/metal contact interfaces. We use the generalized Blonder-Tinkham-Klapwijk (BTK) \cite{Blonder1982,Golubov1995,Perez-Willard2004} model for modeling the superconducting interfaces, identical to the one used for the two-terminal normal metal/graphene/superconductor junctions by Pandey $\textit{et al.}$\,\cite{Pandey2019}. A modified Octavio-Tinkham-Blonder-Klapwijk (OTBK) model \cite{Octavio1983} is employed to address the transport through the NS connections with one superconducting electrode as discussed in Pandey $\textit{et al.}$\,\cite{Pandey2019}.

Using this approach, the transport through the beam splitter, which accounts for the nonlocal processes, can be described by the distribution functions as shown in Fig.\,\ref{fig:CAR_model}(b). Here, $f^{in}$ and $f^{out}$ denote the incoming and outgoing distribution functions in the branches of the beam splitter. For simplicity, we assume that there is no backscattering from the beam splitter into the left superconducting terminal, and that there is no direct transmission between the top and right terminals. The transmissions are constrained by the conservation of probability, and we are left with a single parameter $r=1-\tau_{lr}=\tau_{tl}$. The distribution functions in the beam splitter can be described according to the matrix given below
\begin{equation}
\begin{array}{c}
\begin{bmatrix}
f_\mathrm{l}^{out} \\
f_\mathrm{r}^{out} \\
f_\mathrm{t}^{out} \\
\end{bmatrix}
\hspace{2pt}
=
\hspace{2pt}
\begin{bmatrix}
0 & 1-r & r \\
1-r & r & 0 \\
r & 0 & 1-r \\
\end{bmatrix}
\hspace{2pt}
\begin{bmatrix}
f_\mathrm{l}^{in} \\
f_\mathrm{r}^{in} \\
f_\mathrm{t}^{in} \\
\end{bmatrix}
\end{array}
\label{eqn:dist_func_matrix}
\end{equation} 
Analogous to the OTBK model, the boundary condition for the distribution functions at the superconducting terminal $\alpha$, which is held at a chemical potential $\mu$, can be given as:
\begin{multline}
f_\mathrm{\alpha}^{in}(\epsilon) = T_\mathrm{\alpha}(\epsilon - \mu_\mathrm{\alpha})f_\mathrm{0}(\epsilon - \mu_\mathrm{\alpha}) + R_\mathrm{\alpha}(\epsilon - \mu_\mathrm{\alpha})f_\mathrm{\alpha}^{out}(\epsilon)\\+ A_\mathrm{\alpha}(\epsilon - \mu_\mathrm{\alpha})(1 - f_\mathrm{\alpha}^{out}(-\epsilon + 2\mu_\mathrm{\alpha})).
\label{eqn:dist_func}
\end{multline}
Here $T_{\alpha}$, $R_{\alpha}$ and $A_{\alpha}$ show the normal transmission, normal reflection and Andreev reflection coefficients as obtained from the BTK model. In a similar way, the boundary conditions at the normal terminal $\gamma$ can be expressed by inserting $T_{\gamma} = \tau_{\gamma}$, $R_{\gamma} = 1-\tau_{\gamma}$ and $A_{\gamma} = 0$ in Eq.\,(\ref{eqn:dist_func}). A coupled equation system is obtained by inserting Eq.\,(\ref{eqn:dist_func}) into (\ref{eqn:dist_func_matrix}). Finally, the current in each branch of the beam splitter can be calculated as:
\begin{equation}
I_\mathrm{\alpha} = f_\mathrm{nss}\frac{G_\mathrm{\alpha\beta}}{e}\int (f_\mathrm{\alpha}^{in} - f_\mathrm{\alpha}^{out})\,d\epsilon.
\label{eqn:I_alpha}
\end{equation}
Here $\alpha$ and $\beta$ denote the beam splitter branches, $f_\mathrm{nss}$ is a fit parameter to scale the nonlocal processes relative to the local processes and $G_\mathrm{\alpha\beta}$ gives the two-terminal conductance. While the local processes between the NS and SS terminals are included in the model, they are weighed differently due to the finite transmission through the beam splitter. In addition, the model also includes the nonlocal multiple Andreev reflection (MAR) processes where the MAR cycle between the SS terminals begins due to an initial nonlocal Andreev process.

The model is reduced to a conductance matrix in the normal state as described below:
\begin{equation}
\begin{array}{c}
\begin{bmatrix}
I_\mathrm{l} \\
I_\mathrm{r} \\
I_\mathrm{t} \\
I_\mathrm{b} \\
\end{bmatrix}
\hspace{2pt}
=
\hspace{2pt}
\begin{bmatrix}
\mathcal{G}_\mathrm{ll} & \mathcal{G}_\mathrm{lr} & \mathcal{G}_\mathrm{tl} & \mathcal{G}_\mathrm{bl} \\
\mathcal{G}_\mathrm{lr} & \mathcal{G}_\mathrm{rr} & \mathcal{G}_\mathrm{tr} & \mathcal{G}_\mathrm{br} \\
\mathcal{G}_\mathrm{tl} & \mathcal{G}_\mathrm{tr} & \mathcal{G}_\mathrm{tt} & \mathcal{G}_\mathrm{tb} \\
\mathcal{G}_\mathrm{bl} & \mathcal{G}_\mathrm{br} & \mathcal{G}_\mathrm{tb} & \mathcal{G}_\mathrm{bb} \\
\end{bmatrix}
\end{array}
\hspace{2pt}
\begin{bmatrix}
\mu_\mathrm{l} \\
\mu_\mathrm{r} \\
\mu_\mathrm{t} \\
\mu_\mathrm{b} \\
\end{bmatrix}
\label{eqn:normal_state_matrix}
\end{equation}
where diagonal elements in this matrix are given as:
\begin{equation}
\begin{array}{c}
\mathcal{G}_\mathrm{\alpha\alpha} = -\sum_{\beta\neq\alpha} \mathcal{G}_\mathrm{\alpha\beta}
\end{array}
\label{eqn:cond_sum}
\end{equation}
It should be noted that the conductance parameters $\mathcal{G}_\mathrm{\alpha\beta}$ are not the same as the two-terminal parametric conductance $G_\mathrm{\alpha\beta}$ in Eq.\,(\ref{eqn:I_alpha}). The conductance parameters $\mathcal{G}_\mathrm{\alpha\beta}$ take into account the finite transmission probabilities of the interfaces and additional contribution from the beam splitters. The six independent off-diagonal elements in Eq.\,(\ref{eqn:normal_state_matrix}) can be determined by measuring the six two-probe resistances $\mathrm{R}_\mathrm{\alpha\beta}$ in the normal state. Eq.\,(\ref{eqn:normal_state_matrix}) is then solved for $I_\mathrm{\gamma} = 0$ and $I_\mathrm{\delta} = 0$, where $\mathrm{\gamma}$ and $\mathrm{\delta}$ denote the floating terminals in the two-probe measurement. 

For simplicity, we have kept the ratios $r$ and $f_\mathrm{nss}$ equal for all four beam splitters. We have also set $\tau_\mathrm{l}=\tau_\mathrm{r}=\tau_\mathrm{S}$, and $\tau_\mathrm{t}=\tau_\mathrm{b}=\tau_\mathrm{N}$. Including the conductance parameters $G_{\alpha\beta}$, this leaves us with 10 parameters. An arbitrary set of six parameters can be constrained by the measurements of the six normal-state two-terminal resistances for a given gate voltage. With this constraint, the positions of all the lines in the conductance maps are fixed, and practically independent of the choice of the remaing four free parameters. We have determined the parameters as follows: In the p-doped regime, transport between the superconducting terminals is dominated by the corners (as can be inferred from the normal-state resistances). Therefore, we have set $G_\mathrm{lr}=0$. $\tau_\mathrm{S}$ can be determined quite accurately from the relative weight of the MAR features. In our previous work on similar two-terminal NGS structures, we have found $\tau_\mathrm{N}\sim 0.6-0.7$ in the p-doped regime \cite{Pandey2019}. Since the current maps do not depend much on $\tau_\mathrm{N}$, we have simply chosen $\tau_\mathrm{N}=0.7$. Finally, $r$ was chosen to give a good representation of the small cross-shaped CPS features. The remaining parameters are then constrained. We have also tried to obtain better overall fits using joint least-square fits of several line traces. In addition, we have experimented with lifting the resistrictions outlined above. However, since the weight of the CPS features is quite small, they are not reproduced well by a fit, and the parameters do not change much compared to the ones we chose by hand.

\begin{figure*}[htbp]
\includegraphics[width=1\textwidth]{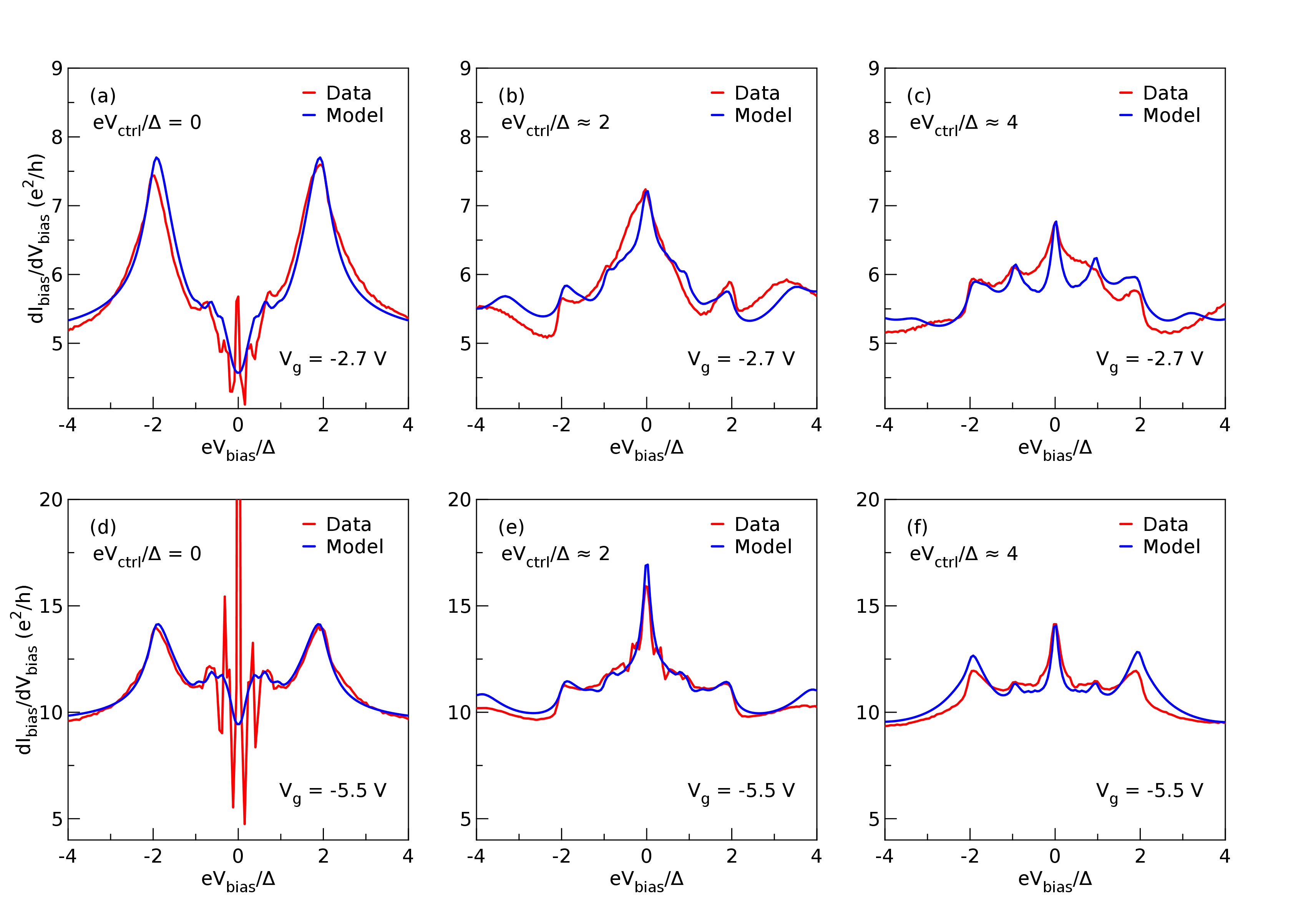}
\caption{Line traces of the measured local differential conductance $dI_\mathrm{bias}/dV_\mathrm{bias}$ in the SGS measurement configuration at $V_\mathrm{g} = -2.7$\,V corresponding to the electronic transport in the vicinity of the SGS Dirac point (panel (a), (b) and (c),) and $T= 20$\,mK, and in the p-doped regime at $V_\mathrm{g} = -5.5$\,V and $T = 50$\,mK (panel (d), (e) and (f)) along with the curves generated from the model at three different $V_\mathrm{ctrl}$ values and zero magnetic field.}
\label{fig:IV_Vg_combined}
\end{figure*}

\begin{table}[htbp]
\centering
\begin{tabular} {ccccccccccc}
\hline\hline
$V_\mathrm{g}$ (V) & $G_\mathrm{lr}$  & $G_\mathrm{tb}$ & $G_\mathrm{tl}$ & $G_\mathrm{tr}$ & $G_\mathrm{bl}$ & $G_\mathrm{br}$ & $\tau_\mathrm{S}$ & $\tau_\mathrm{N}$ & $\mathrm{f_{nss}}$ & $\mathrm{r}$ \\
\hline
-2.7 & $0^{*}$ & 1.22 & 3.67 & 2.38 & 4.57 & 6.09 & $0.72^{*}$ & $0.7^{*}$ & 0.45 & $0.6^{*}$ \\
-5.5 & $0^{*}$ & 17.52 & 5.57 & 4.84 & 4.63 & 5.01 & $0.76^{*}$ & $0.7^{*}$ & 0.73 & $0.47^{*}$ \\
\hline\hline
\end{tabular} 
\caption{Modeling parameters for the data for two different gate voltages. Conductances $G_\mathrm{\alpha\beta}$ are in the units of $e^{2}/h$. Parameters marked with star were chosen to model the experimental data while the others were determined from the two-probe resistances of the device in the normal state.}
\label{table:Model_parameters}
\end{table}

\begin{figure*}
\includegraphics[width=1\textwidth]{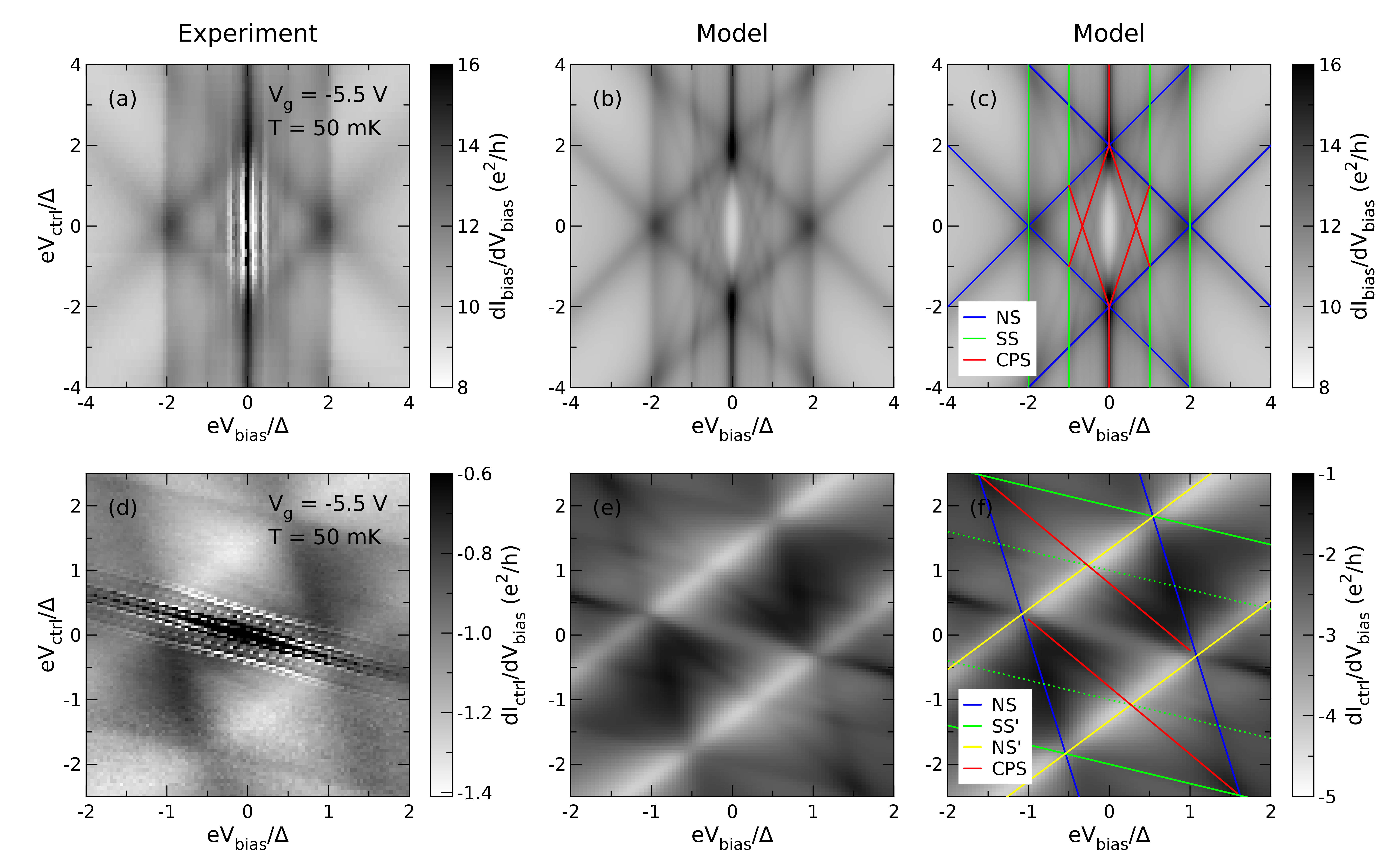}
\caption{Local differential conductance maps, (a) experimentally measured and (b) generated with the model, in the SGS configuration at $V_\mathrm{g} = -5.5$\,V, $T=50$\,mK under zero magnetic field. (c) Same map as shown in (b) but with the guidelines for interpretation. Nonlocal differential conductance maps, (d) experimentally measured and (e) generated with the model, in the NGS configuration under the same measurement conditions as in (a). (f) shows the same map as in (e) with the guidelines.}
\label{fig:data_model_Vg_55}
\end{figure*}

Fig.\,\ref{fig:IV_Vg_combined} shows the line traces of the measured experimental data in the SGS measurement configuration in the vicinity of the Dirac point of the SGS junction at $V_\mathrm{g}= -2.7$\,V and in the p-doped regime at $V_\mathrm{g} =-5.5$\,V along with the curves generated from the model. Note that Fig.\,\ref{fig:IV_Vg_combined}(a), (b) and (c) correspond to the line traces of the data shown in the local differential conductance maps (measured experimentally and generated with the model) in the main text. The modeling parameters for the data are given in Table\,\ref{table:Model_parameters}. It can be readily seen that the model provides a very good fit of the experimental data, except for the oscillatory features at small bias and $V_\mathrm{ctrl}=0$. The sharp peak at zero bias is due to a small supercurrent between the two S terminals which is not accounted for by the model. The sharp oscillatory features at finite bias are attributed to self-induced Shapiro steps. We have also mapped the supercurrent as a function of gate voltage (not shown), and found that the Shapiro features appear at fixed voltage spacings, and generally scale with the magnitude of the critical current. Like the MAR features, the Shapiro features only depend on the bias between the superconducting electrodes, and can therefore not be confused with the CPS features, which also depend on $V_\mathrm{ctrl}$. 

\begin{figure*}
\includegraphics[width=0.9\textwidth]{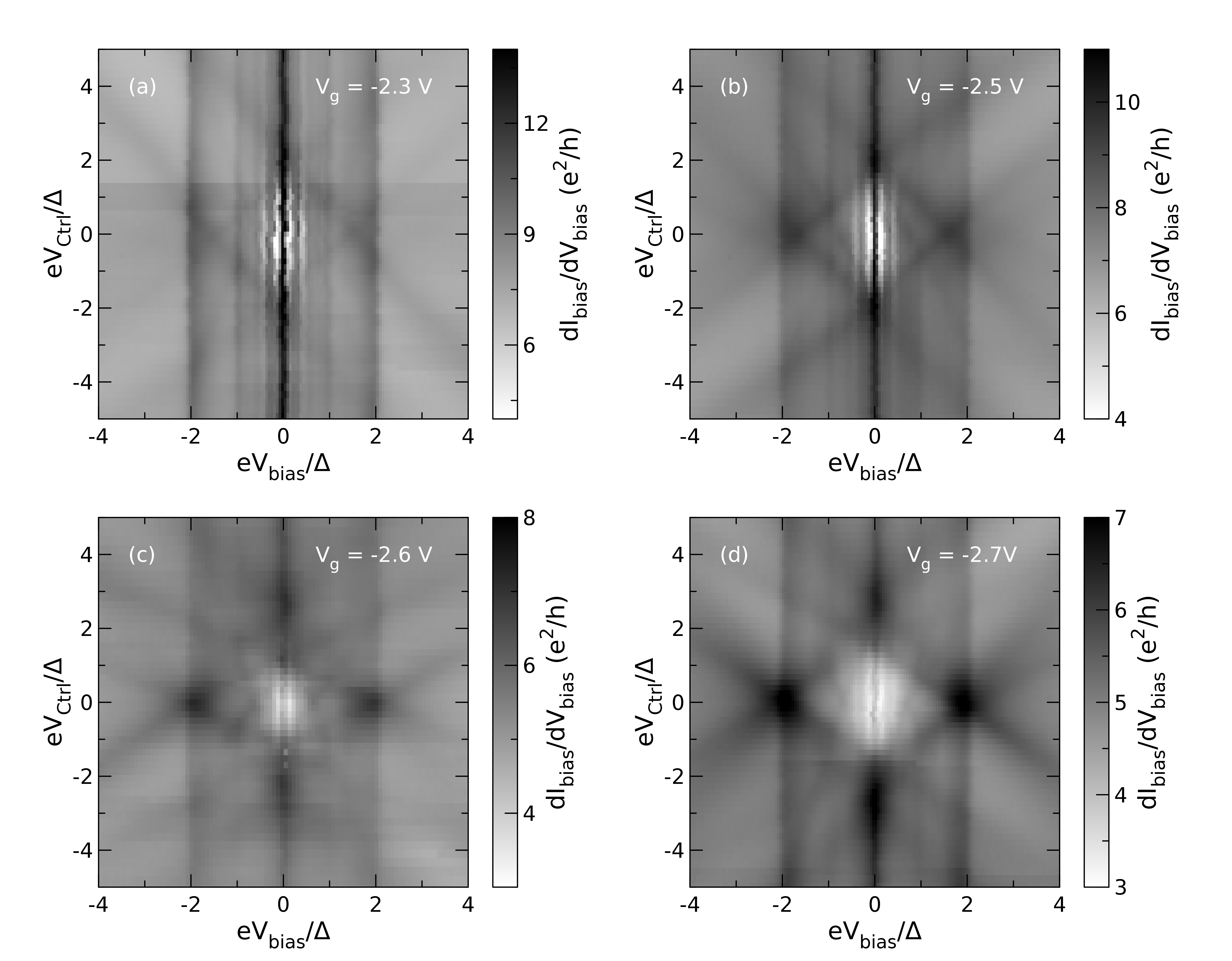}
\caption{Gate dependence of the Cooper pair splitting process close to the charge neutrality point of the SGS junction at (a) $V_\mathrm{g}=$\,-2.3 V, (b) -2.5 V, (c) -2.6 V, (d) -2.7 V.}
\label{fig:CAR_CNP}
\end{figure*}

Fig.\,\ref{fig:data_model_Vg_55}(a) shows the local differential conductance map experimentally measured at $V_\mathrm{g} = -5.5$\,V, $T=50$\,mK under zero magnetic field in the SGS configuration. Fig.\,\ref{fig:data_model_Vg_55} (b) and (c) show the map generated with the model for the same measurement conditions as in (a) without and with the guidelines, respectively. As can be seen by comparing Fig.\,\ref{fig:data_model_Vg_55}(a) and (c), we observe the conductance features due to (i) the electronic transport across the NGS corners of the device (marked NS), (ii) MAR taking place in the SGS junction (marked SS), and (iii) nonlocal transport due to the Cooper pair splitting (marked CPS). Note that the CPS features observed in the experimental data (Fig.\,\ref{fig:data_model_Vg_55}(a)) suffer in terms of visibility due to the larger supercurrent-related features at this gate voltage. However, all of the conductance features can be directly compared with the data shown in the main text for $V_\mathrm{g} = -2.7$\,V which tells us that the observed conductance features are robust and can be very well explained with our model. For completeness, Fig.\,\ref{fig:data_model_Vg_55}(d) shows the nonlocal differential conductance map measured under the same conditions as Fig.\,\ref{fig:data_model_Vg_55}(a) but in the NGS configuration. Fig.\,\ref{fig:data_model_Vg_55}(e) and (f) show the maps generated with the model for the measurement conditions in (d) without and with the guidelines, respectively. Similar to the local differential conductance maps, these nonlocal differential conductance maps can also be easily compared with the data shown in the main text.    

\begin{figure*}
\includegraphics[width=0.9\textwidth]{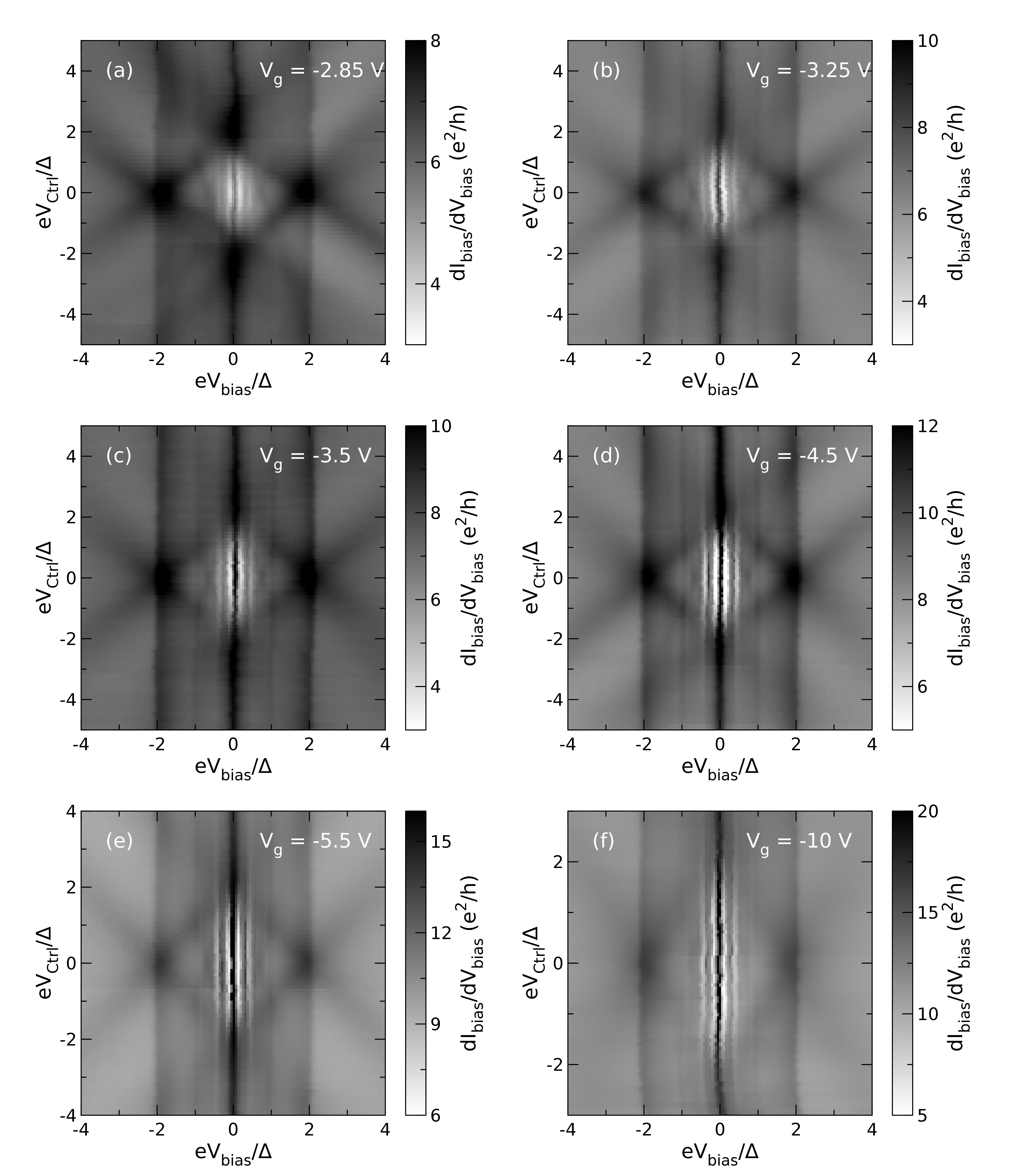}
\caption{Gate dependence of the Cooper pair splitting process in the p-doped regime at (a) $V_\mathrm{g}=$ -2.85 V, (b) -3.25 V, (c) -3.5 V, (d) -4.5 V, (e) -5.5 V, (f) -10 V.}
\label{fig:CAR_h-side}
\end{figure*}

\section{Gate dependence of the Cooper pair splitting}

Here we show the gate dependence of the CPS process in our device. As shown in the main text, the conductance features due to the CPS appear in the low bias regime. It is to be noted that the electronic transport in the low bias regime is mainly tuned by the proximity induced superconductivity in the SGS junction, however, the overall conductance of the graphene channel is tuned by the applied gate voltage. Fig.\,\ref{fig:CAR_CNP} shows the evolution of the conductance features due to the CPS process close to the charge neutrality point (CNP) of the SGS junction. As shown in Fig.\,\ref{fig:CAR_CNP}(a), the SGS junction shows the sign of supercurrent (corresponding to the vertical conductance feature at zero $V_\mathrm{bias}$ in the regime $|eV_\mathrm{Ctrl}/\Delta| < 2$) as well as other conductance features related to the supercurrent (appearing as various vertical conductance features in the low bias regime in the region $|eV_\mathrm{bias}/\Delta| < 1$) at $V_\mathrm{g} = -2.3$\,V (n-doped SGS channel corresponding to an overall higher conductance). Therefore, the conductance features due to the CPS process cannot be clearly distinguished, however, these features become clearly distinguishable as we approach the CNP of the SGS junction as observed in Fig.\,\ref{fig:CAR_CNP}(b), (c) and (d). The role played by the conductance of the graphene channel becomes even clearer as we observe the evolution of the CPS features across the CNP into the p-doped regime shown in Fig.\,\ref{fig:CAR_h-side}. It is clear that the CPS features can be clearly observed in Fig.\,\ref{fig:CAR_h-side}(a) and (b) that correspond to the region close to the CNP. These features can still be observed in Fig.\,\ref{fig:CAR_h-side}(c), (d) and (e), however, they start to suffer in terms of  visibility with the increasing gate voltage, and cannot be distinguished from the background in the high hole-doped regime as seen in Fig.\,\ref{fig:CAR_h-side}(f) at $V_\mathrm{g}=-10$\,V. Note that the data in Fig.\,\ref{fig:CAR_h-side}(f) was collected in smaller $V_\mathrm{bias}$ and $V_\mathrm{Ctrl}$ ranges as compared to the other maps.

\begin{figure*}
\includegraphics[width=1\textwidth]{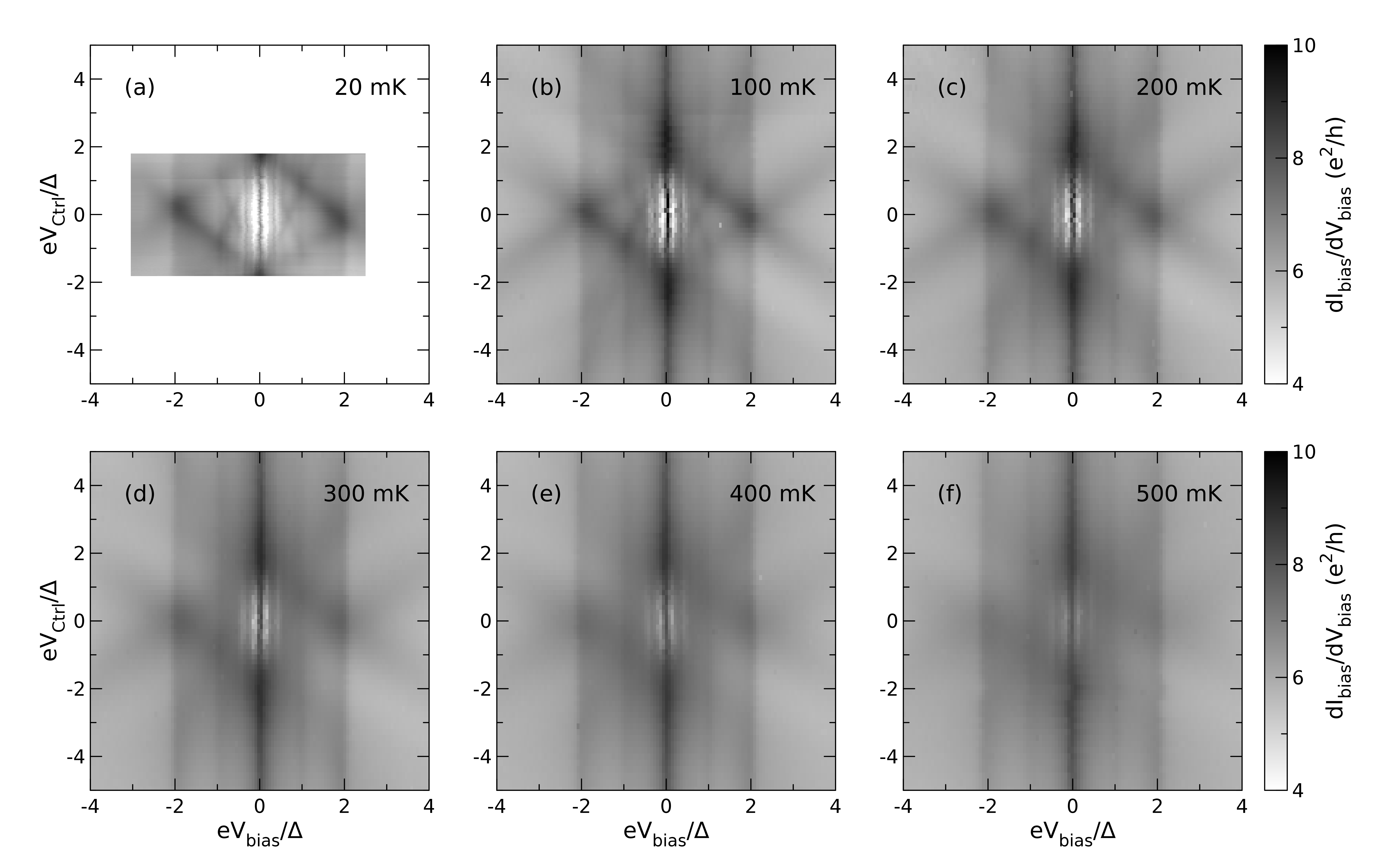}
\caption{Temperature dependence of the Cooper pair splitting process close to the CNP of the SGS junction at $V_\mathrm{g}=-2.68$\,V and under zero magnetic field for (a) $T=20$\,mK, (b) $T=100$\,mK, (c) $T=200$\,mK, (d) $T=300$\,mK, (e) $T=400$\,mK, (f) $T=500$\,mK. Note that the data was collected in a reduced range at $T= 20$\,mK.}
\label{fig:T_dep_CPS}
\end{figure*}

\section{Temperature dependence of the Cooper pair splitting}

Fig.\,\ref{fig:T_dep_CPS} shows the evolution of the CPS conductance features in the vicinity of the CNP of the SGS junction at $V_\mathrm{g}=-2.68$\,V under zero magnetic field with the temperature increasing from 20 mK to 500 mK. Note that the superconductor electrodes used in this device are made of Ti/Al with a superconducting critical temperature $T_\mathrm{C}\sim 900$\,mK. It can be clearly seen that the cross shaped feature in the low bias regime appearing due to the CPS process is clearly observable at 20 mK and 100 mK. While this feature is identifiable as a weak shadow at 200 mK, it becomes completely indistinguishable from the background at 300 mK. This behavior is attributed to the thermal broadening of the gap features resulting in the enhanced background conductance, and hence rendering the cross-shaped features indistinguishable. On the other hand, the vertical conductance ridge appearing at zero bias in the regime $|eV_\mathrm{Ctrl}/\Delta| >2$ is clearly identifiable in the measured temperature range. This feature is robust since the conditions for its appearance are satisfied as long as the device is in the superconducting state (shown in the main text) and it does not suffer from the thermal broadening as severely as the cross shaped feature.